\documentclass[prd,nobibnotes,showpacs,twocolumn]{revtex4}
\usepackage{graphics,color,array,dcolumn}
\usepackage{calc}

\usepackage{amsmath}
\usepackage{amssymb}
\usepackage{xspace}
\newcommand{\eg}{{\it e.g.}}
\newcommand{\ie}{{\it i.e.}}
\newcommand{\tbeta}{\tilde\beta}
\newcommand{\calo}{{\cal O}}
\newcommand{\calm}{{\cal M}}

\newlength{\figurewidth}
\allowdisplaybreaks[1]

\begin{document}
\setlength{\figurewidth}{\columnwidth}

\title{Further Evidence for a Gravitational Fixed Point}
\author{Roberto Percacci}
\email{percacci@sissa.it}
\affiliation{SISSA, via Beirut 4, I-34014 Trieste, Italy}
\affiliation{INFN, Sezione di Trieste, Italy}
\pacs{04.60.-m, 11.10.Hi}
\begin{abstract}
A theory of gravity with a generic action functional and minimally coupled 
to N matter fields has a nontrivial fixed point in the leading large N approximation.
At this fixed point, the cosmological constant and Newton's constant are nonzero 
and UV relevant; the curvature squared terms are asymptotically free 
with marginal behaviour; all higher order terms are irrelevant and can be set to zero 
by a suitable choice of cutoff function.
\end{abstract} \maketitle

There is growing evidence that a generalized version of General Relativity,
possibly including terms of higher order in curvature, may be asymptotically 
safe in the sense of [1].
This property hinges on the existence of a nontrivial Fixed Point (FP) of the
Renormalization Group (RG) flow of the gravitational couplings.
Using Exact Renormalization Group Equation (ERGE) methods,
it has been shown that a FP exists in 4 dimensions within the Einstein-Hilbert truncation
(where only the cosmological constant and Newton's constant are retained)
and there are various arguments suggesting that this FP is not a mere
artifact of the truncation [2,3].
Independent evidence for the existence of a FP comes from recent results
in causal dynamical triangulations [4].
Since in the real world gravity does not exist in isolation, but rather
interacts with many matter fields, it is important to establish that
matter interactions do not spoil the FP.
(In fact, for applications to physics it is not logically required that pure 
gravity exists as a quantum theory: it is only necessary that a theory of gravity and
realistic matter exists).
In this direction, positive evidence has come from the analysis
of arbitrary matter fields minimally coupled to gravity in the Einstein-Hilbert
truncation [5], and from the study of the
coupled scalar-gravity theory with arbitrary couplings of the form
$\phi^{2n}$ and $\phi^{2n}R$ with $n\geq 0$ (thus including the original
Einstein-Hilbert terms) [6].

One of the central issues of this approach 
is the extension of the results to include higher gravitational couplings.
The beta functions of four--derivative gravity have been the subject
of several papers in the past [7], also [8] and references therein; 
for the state of the art on this subject see [9].
Analyzing the structure of divergences in dimensional regularization,
it has been shown that the couplings of the terms quadratic in curvature
are asymptotically free;
however, due to the fact that dimensional regularization is not
well--suited to the discussion of mass and threshold effects, the
behaviour of the cosmological constant and Newton's constant
has not been thoroughly understood.
In [10] a term proportional to $R^2$ was included in the truncation of the ERGE
and it was shown that its presence does not spoil the existence of the FP.
In fact, the FP values of the cosmological constant and Newton's constant
are only slightly changed relative to the Einstein--Hilbert truncation, 
and the FP--value of the new term is quite small.

In this note we shall consider a different approximation scheme,
namely the large $N$ expansion, where $N$ is the number of matter fields.
This approximation has been first applied to gravity in [11] and 
was later used in [12] to prove the existence of a FP.
I will use the ERGE in conjunction with the large $N$ 
approximation and the heat kernel expansion to rederive and extend this result.
The truncation will consist of assuming that all matter self-interactions
can be neglected. 
Nothing will be assumed a priori about the gravitational action.
In the leading order in $N$ the analysis is so simplified, 
that we will be able to prove the existence of the FP for arbitrary 
gravitational couplings, and furthermore 
we shall prove that there exists a scheme (\ie\ a choice
of cutoff function) such that all the coefficients of terms of cubic
and higher order in the curvature vanish.

The ERGE can be written, slightly symbolically,
\begin{equation}
\label{eq:erge}
\partial_t \Gamma_k=
{1\over2}{\rm Tr}\left({\delta^2\Gamma_k\over\delta \Phi\delta \Phi}+R_k\right)^{-1}\partial_tR_k
\end{equation}
where $\Gamma_k$ is the coarse--grained effective action,
$\Phi$ are all the fields in the theory, Tr denotes a trace over all
degrees of freedom of the theory (sum over indices and integration over momenta),
$R_k$ is a cutoff function that suppresses the propagation of modes with
momenta lower than $k$ and $t=\log k$. The RG flow of the theory can be computed by making
an Ansatz for the effective action $\Gamma_k$, inserting it in (\ref{eq:erge})
and reading off the beta functions of the couplings.

For simplicity we will begin by considering gravity coupled to $N$ real scalar fields.
In the spirit of an effective field theory, we will consider arbitrary
gravitational actions of the form
\begin{equation}
\label{eq:action}
S_{grav}(g_{\mu\nu};g^{(n)}_i)=\sum_{n=0}^\infty\sum_i g^{(n)}_i \calo^{(n)}_i
\end{equation}
where $\calo^{(n)}_i=\int d^nx\,\sqrt{g}\calm^{(n)}_i$ 
and $\calm^{(n)}_i$ are polynomials in the curvature and its derivatives
involving $2n$ derivatives of the metric. 
The index $i$ is used to label different polynomials having the same $n$
(and hence the same mass dimension $2n$).
The first two polynomials are simply
$\calm^{(0)}=1$ and $\calm^{(1)}=R$.
The corresponding couplings are
$g^{(0)}=2 Z_g\Lambda$, where $\Lambda$ is the cosmological constant and
$g^{(1)}=-Z_g=-{1\over16\pi G}$, where $G$ is Newton's constant.
For $n=2$ I will follow [7-9] and choose as independent polynomials
\begin{equation}
\begin{aligned}
\calm^{(2)}_1=&C_{\mu\nu\rho\sigma}C^{\mu\nu\rho\sigma}=
R_{\mu\nu\rho\sigma}R^{\mu\nu\rho\sigma}-2R_{\mu\nu}R^{\mu\nu}+{1\over3}R^2\\
\calm^{(2)}_2=&G=
R_{\mu\nu\rho\sigma}R^{\mu\nu\rho\sigma}-4R_{\mu\nu}R^{\mu\nu}+R^2\\
\calm^{(2)}_3=& R^2\  ;\ \ \calm^{(2)}_4=\nabla^2 R\ ,\\
\end{aligned}
\end{equation}
where $C$ is the Weyl tensor and $G$ is $32\pi^2$ times the integrand of the Gauss--Bonnet
topological invariant. 
The coefficients $g^{(2)}_1$ and $g^{(2)}_3$ are sometimes denoted
$1/\lambda$ and $-\omega/3\lambda$ [8,9].
The second and fourth terms are total derivatives and do not
contribute to local physical processes.
We will assume that the scalar fields are all massless and minimally coupled
to gravity,  with inverse propagator $z=-\nabla^2$.
For the running effective action $\Gamma_k$ we assume that it is the sum of the
scalar action and (\ref{eq:action}), 
where all couplings $g^{(n)}_i$ are supposed to be $k$-dependent.
This defines our truncation.
We will discuss later the stability of this truncation against quantum fluctuations.

In the $1/N$ expansion one assumes that the number of matter fields
$N\to\infty$ [11]. 
Here we shall restrict ourselves to the leading order approximation, where one simply
neglects the gravitational contribution with respect to the matter contribution.
This eliminates most of the complications that arise in the gravitational 
ERGE calculations and provides the simplest way of arriving at a gravitational FP.
Taking into account the cutoff function, 
the modified inverse propagator is $P_k(z)=z+R_k(z)$;
we can then write the ERGE (\ref{eq:erge})
\begin{equation}
\partial_t\Gamma_k=
{N\over2}{\rm Tr}\left({\partial_t P_k\over P_k}\right) 
\end{equation}

The trace of a function $f$ of the Laplacian can be written as
\begin{equation}
{\rm Tr}f\equiv\sum_i f(\lambda_i)=\int_0^\infty dt {\rm Tr}K(t)\tilde f(t)
\end{equation}
where $\lambda_i$ are the eigenvalues of the Laplacian,
$\tilde f$ is the Laplace transform of $f$
and ${\rm Tr}K(t)=\sum_n e^{-t\lambda_n}$ is the trace of the heat kernel of $-\nabla^2$.
(We assume that there are no negative and zero eigenvalues; if present, these will
have to be dealt with separately.)

One can then use the known properties of the heat kernel to calculate the trace
in various circumstances.
If we are interested in the local behaviour of the theory (\ie\ the behaviour
at scales $k$ much smaller than the typical curvature) 
we can use the asymptotic expansion
\begin{equation}
\label{eq:asymptotic}
{\rm Tr}K(t)\approx B_0t^{-2}+B_2t^{-1}+B_4+B_6t+\ldots
\end{equation}
where $B_{2k}=\int\,d^4x\,\sqrt{g}\, b_{2k}$, $b_{2k}$ being well-polynomials 
in the curvature tensor and its covariant derivatives. Then we get
\begin{equation}
{\rm Tr}f=B_0Q_2(f)+B_2Q_1(f)+B_4Q_0(f)+B_6Q_{-1}(f)+\ldots
\label{eq:trace}
\end{equation}
where 
\begin{equation}
Q_n(f)=\int_0^\infty dt t^{-n}\tilde f(t)
\end{equation}
We can rewrite these integrals in terms of the original function $f$:
for $n\geq 1$ it is a Mellin transform 
\begin{equation}
Q_n(f)={1\over\Gamma(n)}\int_0^\infty dz z^{n-1} f(z)
\end{equation}
whereas for $n\leq 0$
\begin{equation}
Q_n(f)=(-1)^n f^{(n)}(0)
\end{equation}
where $f^{(n)}$ denotes the $n$-th derivative of $f$.
Using (\ref{eq:trace}),
\begin{widetext}
\begin{equation}
\begin{aligned}
\label{eq:scalars}
\partial_t\Gamma_k=
{N\over2}{1\over(4\pi)^2}\int\,d^4x\,\sqrt{g}
\Bigg[
&
Q_2\left({\partial_t P_k\over P_k}\right) +
{1\over 6}R Q_1\left({\partial_t P_k\over P_k}\right)\\
&+
{1\over180}\left(\frac{3}{2}C^2-{1\over2}G +\frac{5}{2} R^2+6\nabla^2R\right)
Q_0\left({\partial_t P_k\over P_k}\right)
+\ldots
\Biggr]\\
\end{aligned}
\end{equation}
\end{widetext}
Define 
$\partial_t g^{(n)}_i=\beta^{(n)}_i=k^{4-2n}a^{(n)}_i$.
The coefficient of each term in (\ref{eq:scalars})
is the beta function of the corresponding coupling.
The beta functions of the dimensionless couplings
$\tilde g^{(n)}_i= g^{(n)}_i k^{2n-4}$ are given by
\begin{equation}
\partial_t \tilde g^{(n)}_i=\tbeta^{(n)}_i=(2n-4)\tilde g^{(n)}_i+a^{(n)}_i
\label{eq:betafunctions}
\end{equation}
where the numbers $a^{(n)}_i$ can be read off (\ref{eq:scalars}).
For all $n\not= 2$ this leads to a fixed point
\begin{equation}
\tilde g^{(n)}_{i*}={1\over 4-2n}a^{(n)}_i
\end{equation}
For $n=2$ one gets instead the following solution:
\begin{equation}
g^{(2)}_i(k)=g^{(2)}_i(k_0)+a^{(2)}_i\ln(k/k_0)
\end{equation}
As noted in [1], the condition of reaching a FP has to be
imposed on those parameters for which a divergence 
would lead to divergent physical reaction rates.
In the Wilsonian approach used here, it is enough to 
consider the tree level formulas deduced from the effective action.
In the case of the terms $\calm^{(2)}_1$ and $\calm^{(2)}_3$, 
the interactions are not proportional $a^{(2)}_1$ and $a^{(2)}_3$, 
but rather to their inverses (in the notation of [8.9],
they are given by $\lambda$ and $-3\lambda/\omega$).
The logarithmic running of the $a^{(2)}$'s implies a FP,
more specifically, asymptotic freedom, for these coupling constants.
The physical significance of the parameters $a^{(2)}_2$ and $a^{(2)}_4$,
which are the coefficients of total derivative terms,
is less clear in this respect, and we simply observe that the
running derived above implies asymptotic freedom for their inverses.

Choosing a cutoff function $R_k$ determines the numbers $a^{(n)}_i$ and hence affects the
position of the FP. This choice does not affect physically measurable 
quantities, however.
One particularly convenient choice of cutoff function is the so-called optimized cutoff
$R_k(z)=(k^2-z)\theta(k^2-z)$, where $\theta$ is the Heaviside step function [13].
With this choice we get
$Q_2\left({\partial_t P_k\over P_k}\right)=k^4$,
$Q_1\left({\partial_t P_k\over P_k}\right)=2k^2$,
$Q_0\left({\partial_t P_k\over P_k}\right)=2$ and
$Q_n\left({\partial_t P_k\over P_k}\right)=0$ for $n\leq -1$.
In terms of the cosmological constant and Newton's constant, the FP occurs for 
\begin{equation}
\tilde\Lambda=\Lambda k^{-2}=-\frac{3}{4}\ ;\qquad\qquad
\tilde G=G k^2=-{12\pi\over N}
\end{equation}
The sign of Newton's constant is unphysical, but this is
only a feature of the scalar fields (see eq.(18)).
Perhaps the biggest surprise is the fact that in this scheme the FP--value of all the couplings with $n \geq 3$ is zero. This is due to the flatness of the function ${\partial_t P_k\over P_k}$
for small $z$.
With the optimized cutoff, it would therefore be consistent to neglect all the terms
with more than four derivatives of the metric.

The dimension of the critical surface is determined by the linearized flow,
which is defined by the matrix $M_{ij}={\partial\tbeta_i\over\partial\tilde g_j}\big|_*$.
Since the $a^{(n)}_i$ in (\ref{eq:betafunctions}) 
are independent of the couplings, the eigenvalues of $M$
are equal to the canonical dimensions.
Therefore, the cosmological constant and Newton's constant are UV relevant (attractive),
the $g^{(2)}_i$ couplings are marginal and all the higher terms are irrelevant.

With the same techniques, one can explore also other regimes.
For large $t$, the heat kernel on a noncompact manifold has
the asymptotic behaviour $K(t)\approx {1\over (4\pi)^2}t^{-1} S$ where [15]
$$
S= \int d^4x\,\sqrt{g}
\left[\frac{1}{6}R+\frac{1}{12}R\frac{1}{\nabla^2}R
-\frac{2}{3}R_{\mu\nu}\frac{1}{\nabla^2}R^{\mu\nu}+\ldots\right]
$$
Since all the terms have the same dimension, they also have the
same overall running.
The beta functions of the operators appearing in $S$
are equal to ${1\over (4\pi)^2}Q_1(\frac{\partial_t P_k}{P_k})$, 
times the coefficient of the operator as it appears in $S$.
Consequently, there is a FP where the FP--action is proportional to $S$,
the exact coefficient being scheme--dependent.
With the optimized cutoff, the FP--action is $\Gamma_*=\frac{N}{64\pi^2}k^2 S$.
We note that nonlocal actions of this form may have some relevance
to cosmology [16].

Let us now consider the effect of other matter fields.
Following [14], we can calculate the contribution to the ERGE of $n_S$ scalar fields,
$n_D$ Dirac fields, $n_M$ Maxwell fields, all massless and minimally coupled.
Unlike in [14], however, we shall not make any assumption about the background metric,
so as to be able to discriminate the coefficients of various contractions
of two curvatures.
For each type of field we choose the cutoff function in such a way that
the modified propagator has the form $P_k(\calo)$, where
$\calo^{(S)}=-\nabla^2$ on scalars, $\calo^{(D)}=-\nabla^2+\frac{R}{4}$  on Dirac fields 
and $\calo^{(V)}=-\nabla^2\delta^\mu_\nu+R^\mu{}_\nu$ 
on Maxwell fields in the gauge $\lambda=1$.
We can then use the heat kernel coefficients given \eg\ in [17]. 
In the case of the Maxwell fields,
the calculation of the effective action with nonminimal operators 
(corrisponding to other gauges) was given in [18], 
and shows that the results are gauge--independent.

We assume that the numbers $n_S$, $n_W$ and $n_M$ are all of order $N$, or zero.
The ERGE becomes
\begin{equation}
\begin{aligned}
\partial_t\Gamma_k=&
{n_S\over2}{\rm Tr}_{(S)}\left({\partial_t P_k\over P_k}\right) -
{n_D\over2}{\rm Tr}_{(D)}\left({\partial_t P_k\over P_k}\right) +\\
&+{n_M\over2}{\rm Tr}_{(M)}\left({\partial_t P_k\over P_k}\right) -
{n_M}{\rm Tr}_{(S)}\left({\partial_t P_k\over P_k}\right) \\
\end{aligned}
\end{equation}
where the argument of each function appearing under the traces
is the appropriate operator $\calo$ and the last term is due to the ghosts.
With the optimized cutoffs,
the beta functions are characterized by the following coefficients:
\begin{equation}
\begin{aligned}
a^{(0)}=&\frac{1}{32\pi^2}\left(n_S-4n_D+2n_M\right)\\
a^{(1)}=&\frac{1}{96\pi^2}\left(n_S+2n_D-4n_M\right)\\
a^{(2)}_1=&\frac{1}{2880\pi^2}\left(\frac{3}{2}n_S+9 n_D+18n_M\right)\\
a^{(2)}_2=&\frac{1}{2880\pi^2}\left(-\frac{1}{2}n_S-\frac{11}{2}n_D-31n_M\right)\\
a^{(2)}_3=&\frac{1}{2880\pi^2}\frac{5}{2}n_S\\
a^{(2)}_4=&\frac{1}{2880\pi^2}\left(6n_S+6n_D-18n_M\right)\\
\end{aligned}
\end{equation}
The fixed point occurs at
\begin{equation}
\begin{aligned}
\tilde\Lambda_*=&-\frac{3}{4}\,{n_S-4n_D+2 n_M\over n_S+2n_D-4n_M}\\
\tilde G_*=&{12\pi\over -n_S-2n_D+4n_M}\\
\end{aligned}
\end{equation}
We see that there is a FP for any value of $n_S$, $n_W$ and $n_M$.
As expected, bosons and fermions contribute with opposite signs to the cosmological constant.
Newton's constant is only positive if the ratio of scalar and Dirac to Maxwell fields 
is not too large. 
The couplings $(g^{(2)}_i)^{-1}$ are asymptotically free,
tending to zero from above or below depending on the sign of $a^{(2)}_i$,
and again $g^{(n)}_{i*}=0$ for $n\geq 3$.
The sign of $a^{(2)}_1$ and $a^{(2)}_3$ guarantees that the Euclidean path integral
is damped.
Since this action represents the asymptotic UV behaviour of the theory,
it is not required to reproduce the observed large--scale behaviour of gravity.

Let us now discuss the consistency of the truncation
that has been used in this paper.
In general, one could expect processes with intermediate gravitons to generate
matter interactions even if they were initially set to zero.
However, this will not happen in the UV limit if the interactions are asymptotically free.
It has been shown in [6] that a large class of scalar interactions
becomes asymptotically free in the presence of gravity.
Thus, the truncation we used here for the scalar fields is probably consistent.
Gauge fields are asymptotically free already in flat space and it
is reasonable to expect that the situation will not change
in the presence of dynamical gravity.
More troublesome are the Yukawa couplings, which in the standard 
model have positive beta functions.
It is tempting to conjecture that the phenomenon that occurs for
scalar fields is more general, and that there exists a ``Gaussian--Matter FP''
where all the matter interactions are asymptotically free and only
the purely gravitational couplings (\ref{eq:action}) have novanishing values at the FP.
Earlier calculations reported in [8] suggest that this is the case.
Then, the truncation adopted in this paper with regard to the matter fields
would be a consistent truncation.

The use of the large $N$ limit has several advantages over previous calculations.
First of all, it dramatically simplifies the derivation of the beta functions,
making the result very trasparent.
Furthermore, the calculations presented here are free of the gauge fixing 
ambiguities that occur when graviton loops are taken into account. 
This shows that the FP cannot be an artifact of the gauge choice.
The most important result derived here is the existence of the
FP for all the terms of higher order in curvature.
What is perhaps even more striking, 
all the couplings with $n\geq 3$ can be made to vanish
by a simple choice of cutoff scheme.
It is known that the position of the FP is scheme dependent, 
but in the case of the cosmological constant and
Newton's constant no sensible scheme is known were they vanish at the FP.

One of the problems with applications of the ERGE is the absence of systematic
ways of estimating the errors due to the truncation. 
If one can work with a sufficiently general consistent truncation,
the large $N$ limit introduces just such an expansion parameter.
Calculating the next--to--leading order in $1/N$ requires the evaluation
of graviton contributions. 
This is obviously impossible with the full gravitational action
(\ref{eq:action}), but the results presented here suggest that 
in certain schemes it may be consistent to set to zero all terms with $n\geq 3$.
This would not be an {\it a priori}, uncontrollable truncation:
due to the scheme--independence of the physical results,
it would be equivalent to calculating with the full action.

\centerline{\bf Acknowlegements}
I wish to thank M. Reuter for discussions and correspondence.

\centerline{\bf References}

[1] S. Weinberg, 
In {\it General Relativity: An Einstein centenary survey}, 
ed. S.~W. Hawking and W. Israel, chapter 16, pp.790--831; 
Cambridge University Press (1979).

[2] W. Souma, Prog. Theor. Phys. {\bf 102}, 181 (1999),
[arXiv:hep-th/9907027].

[3] O. Lauscher and M. Reuter, {\it Phys. Rev.} {\bf D65}, 025013 (2002),
[arXiv:hep-th/0108040]; 
Class. Quant. Grav. {\bf 19}, 483 (2002),
[arXiv:hep-th/0110021]; 
Int. J. Mod. Phys. {\bf A 17}, 993 (2002).
[arXiv:hep-th/0112089];
M. Reuter and F. Saueressig, Phys. Rev. {\bf D65}, 065016 (2002),
[arXiv:hep-th/0110054].

[4] J. Ambj\o rn, J. Jurkiewicz, R. Loll, 
Phys. Rev. Lett. {\bf 95} 171301 (2005),
[arxiv:hep-th/0505113]; 
Phys. Rev. {\bf D72} 064014 (2005),
[arxiv:hep-th/0505154].

[5] R. Percacci and D. Perini, {\it Phys. Rev.} {\bf D67}, 081503(R) (2003),
[arXiv:hep-th/0207033].

[6] R. Percacci and D. Perini, {\it Phys. Rev.} {\bf D68}, 044018 (2003),
[arXiv:hep-th/0304222].

[7] E.S. Fradkin, A.A. Tseytlin, 
Phys. Lett. {\bf 104 B}, 377 (1981);
Nucl. Phys. {\bf B 201}, 469 (1982);
I.G. Avramidi, A.O. Barvinski, 
Phys. Lett. {\bf 159 B}, 269 (1985).

[8] I.L. Buchbinder, S.D. Odintsov and I.L. Shapiro, 
``Effective action in quantum gravity'',
IOPP Publishing, Bristol (1992).

[9] de Berredo--Peixoto and I. Shapiro, Phys.Rev. {\bf D71} 064005 (2005)
[arXiv:hep-th/0412249]

[10] O. Lauscher and M. Reuter, {\it Phys. Rev.} {\bf d 66}, 025026 (2002).

[11] E.T. Tomboulis, Phys. Lett. {\bf 70 B} 361 (1977),
[arXiv:hep-th/9601082]; Phys. Lett. {\bf 97 B} 77 (1980).

[12] L. Smolin, Nuclear Physics {\bf B 208}, 439-466 (1982).

[13] D.F. Litim, Phys.Rev. {\bf D 64} 105007 (2001),
[arXiv:hep-th/0103195]; 
Phys.Rev.Lett. {\bf 92} 201301 (2004),
[arXiv:hep-th/0312114].

[14] D. Dou and R. Percacci, Class and Quantum Grav. {\bf 15}, 3449 (1998).

[15] A.O. Barvinski and G.A. Vilkoviski, {\it Nuclear Physics} {\bf B 333}, 471 (1990).

[16] C. Wetterich, Gen. Rel. and Grav. {\bf 30}, 159-172 (1998),
[arXiv: gr-qc/9704052].

[17] S.M. Christensen and M.J. Duff, Nucl. Phys. {\bf B 154}, 301 (1979).

[18] A.O. Barvinski and G.A. Vilkoviski, Phys. Rep. {\bf 119}, 1 (1985).
\end{document}